\documentclass[3p,authoryear]{elsarticle}
\usepackage{times,graphics,amssymb,amsmath,epsfig,subfigure,relsize}
\journal{New Astronomy}
\usepackage{aas_macros,hyperref}

\newcommand{\bea}{\begin{eqnarray}}
\newcommand{\eea}{\end{eqnarray}}

\begin{document}
\begin{frontmatter}

\title
{Photo-$z$ with CuBAN$z$: An improved photometric redshift estimator using
Clustering aided Back Propagation Neural network}
\author{Saumyadip Samui\corref{cor1}}
\ead{ssamui@gmail.com}
\author{Shanoli Samui Pal\corref{cor2}}
\ead{shanoli.pal@gmail.com}
\address[cor1]{Department of Physics, Presidency University, 86/1 College
Street, Kolkata - 700073, India.}
\address[cor2]{Department of Mathematics, NIT Durgapur, Durgapur 713209, India.}
\cortext[cor1]{Department of Physics, Presidency University, 86/1 College
Street, Kolkata - 700073, India.}
\cortext[cor2]{Department of Mathematics, NIT Durgapur, Durgapur 713209, India.}

\begin{abstract}
We present an improved photometric redshift estimator code, CuBAN$z$, that is
publicly available at \href{https://goo.gl/fpk90V}{https://goo.gl/fpk90V}. It uses the back propagation
neural network along with clustering of the training set, which makes it
more efficient than existing neural network codes. In CuBAN$z$, the training
set is divided into several self learning clusters with galaxies having
similar photometric properties and spectroscopic redshifts within a given span.
The clustering algorithm uses the color information (i.e. $u-g$, $g-r$ etc.)
rather than the apparent magnitudes at various photometric bands as the photometric
redshift is more sensitive to the flux differences between different bands
rather than the actual values. Separate neural networks are trained for each
cluster using all possible colors, magnitudes and uncertainties in the
measurements. For a galaxy with unknown redshift, we identify the closest
possible clusters having similar photometric properties and use those clusters
to get the photometric redshifts using the particular networks that were trained
using those cluster members. For galaxies that do not match with any training
cluster, the photometric redshifts are obtained from a separate network
that uses entire training set. This clustering method enables us to
determine the redshifts more accurately. SDSS Stripe 82 catalog
has been used here for the demonstration of the code. For the clustered sources
with redshift range $z_{\rm spec}<0.7$,
the residual error ($\langle (z_{{\rm spec}}-z_{{\rm phot}})^2\rangle^{1/2} $)
in the training/testing phase is as low as 0.03 compared
to the existing ANNz code that provides residual error on the same test
data set of 0.05. Further, we provide a much better estimate of the
uncertainty of the derived photometric redshift.
\end{abstract}
\end{frontmatter}

\section{Introduction}
Even though there is a huge advancement in the telescope technology,
spectroscopy of a large number of galaxies is still very time expensive
especially for high redshift large scale galaxy surveys. Thus photometry
is still the best bet for such surveys whether they are the existing
ones, i.e. Solan Digital Sky Surveys (SDSS), 2dF Galaxy
redshift Survey, Blanco Cosmological Survey, Dark Energy Survey
\citep{2014ApJS..211...17A,1999RSPTA.357..105C,2015ApJS..216...20B}
or the future
planed ones like Large Synoptic Survey Telescope \citep{2008arXiv0805.2366I}, etc.
Hence we need to infer redshift of the sources from the photometric
measurements only. Two types of photometric redshift (photo-$z$)
determination processes are vastly used. One is the template base analysis
such as HyperZ, ImpZ, BPZ, ZEBRA
\citep{2000A&A...363..476B,2004MNRAS.353..654B,2000ApJ...536..571B,2006MNRAS.372..565F}.
The other uses the neural networks
to get empirical relation between redshift and available colors, such as
ANNz, ArborZ
\citep{2004PASP..116..345C,2010ApJ...715..823G}. Recently,
some other techniques have also been proposed to get the photo-$z$ such
as genetic algorithm, gaussian processes etc., \citep{2015MNRAS.449.2040H, 2010MNRAS.405..987B}.
Both template fitting and neural network
approaches possess their merits and demerits \citep{2011MNRAS.417.1891A}.
The template base redshift determinations
are always biased from the available templates and need to know
the filter response, detector response etc., very well.
On the other hand, the neural network methods provide
better results than the template analysis method
if there are large number of galaxies available for the training
set. Given the present day increase in the number of spectroscopic
sample of galaxies, this method would be the best possible
choice and thus it's timely to make some improvement
on it.

Here, we propose an improved technique that uses existing neural
network algorithm combined with clustering of the training set galaxies
in order to get more accurate photometric redshifts for sources
with known photometric properties. Our
method is better in the following ways. We use a back propagation of
error to train the neural networks compare to the existing
ANNz code that uses quasi-Newton method \citep{2004PASP..116..345C}.
Secondly and most importantly, we build self-learning clusters
from the training set with galaxies having similar photometric
properties and spectroscopic redshifts. Our modified clustering algorithm
takes into account of the uncertainty in the measurements where as 
the traditional clustering algorithms just ignore these uncertainties.
Separate neural networks are trained using the members of
each clusters. The training of neural networks are done
considering all possible differences in photometric
magnitudes between different bands (i.e. the color) along with the
apparent magnitudes in each bands and the errors associated with them.
It allows us to map the redshift from the photometric measurements
more accurately as colors are more sensitive to redshift.
In order to obtain photometric redshift of unknown
sources, we first seek for clusters that have similar
photometric properties. If there is any, the neural networks that
are trained using those clusters are used to find the photometric
redshift of that galaxy. Otherwise, a separate network which is trained using
all available galaxies for the training
is used to get the photometric redshift.
This ensures a much more accurate estimate of the redshift for the sources that
match with clusters having similar properties in the training set.
Finally, we provide more realistic treatment to estimate
the uncertainty in the derived photometric redshift
by considering the possible uncertainty in the training process,  
so that it can be used more confidently for further analysis of the galaxy properties
such as number distribution, finding groups/clusters of galaxies etc.

The paper is organised as follows. In section~\ref{sec_data} the
data set that has been used in this paper is described in details.
Our clustering models are discussed in 
section~\ref{sec_cluster}. The back propagation neural network is
described in section~\ref{sec_bnn}. We show the performance of our
code in section~\ref{sec_result}.
In section~\ref{sec_code} we describe
our code and its usages. Finally we discuss and conclude in
section~\ref{sec_cd}.

\section{The data set}
\label{sec_data}
\begin{figure}
\centerline{\epsfig{figure=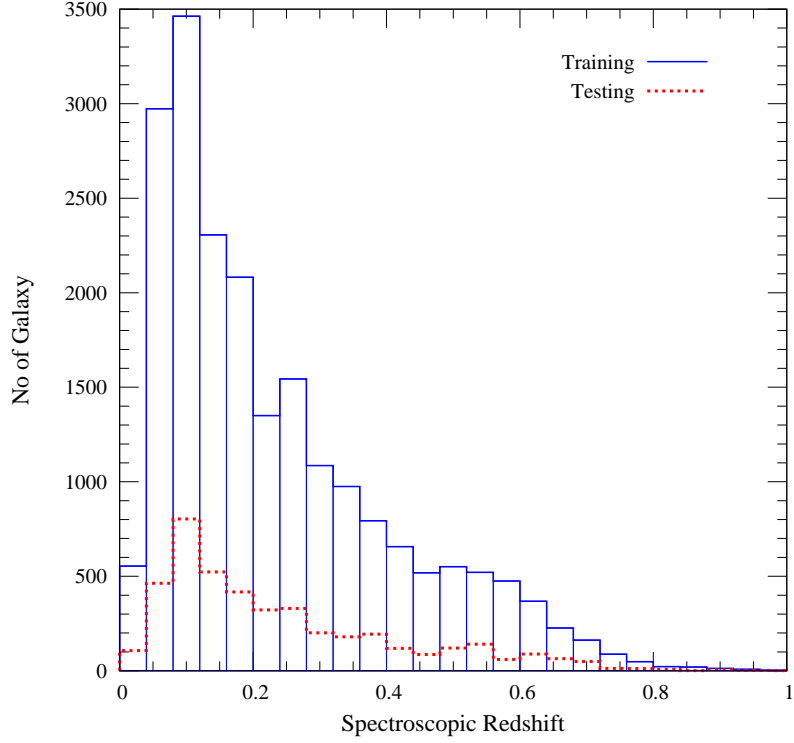,width=10cm,angle=-90.}}
\caption[]{Redshift distribution of galaxies in the training set
(blue solid histogram) and the testing set (red dotted histogram) for CuBAN$z$.}
\label{fig_dist_test_train}
\end{figure}

SDSS stripe 82 catalog has been used for the analysis of the present work
\citep{2014ApJ...794..120A}.
SDSS stripe 82 catalog has an area of 270 square
degrees and is two magnitudes fainter than normal SDSS catalog. Galaxies
have 5 bands photometry namely, $u$, $g$, $r$, $i$ and $z$. We take
the entire `galaxy' catalog and search for availability
of spectroscopic redshift in SDSS. We find total 25120 galaxies having
spectroscopic redshifts. Out of these, we construct the training set consisting
of 20809 galaxies and a separate testing set of 4311 galaxies.
The spectroscopic redshift distributions for both training data set
and validation data set are shown in Fig.~\ref{fig_dist_test_train}.
Note that same data set is used to show results considering only 4 bands
removing the $z$ band photometry.
Further, we use our code on the entire stripe 82 catalog to find photometric
 redshift. In this case we use galaxies with $r<23.26$ that is 50\%
completeness limit of the stripe 82 catalog, along with cut off on the
photometric uncertainty of $|\delta m| < 0.2$ in $g$, $r$ and $i$ bands
\citep{2014ApJ...794..120A}.

\section{Clustering algorithm of sources}
\label{sec_cluster}
The clustering of data is a widely used statistical tool that has
applications in various fields such as machine learning, optimisation,
forecasting etc. Different clustering mechanism includes hierarchical clustering,
density based, centroid based clustering etc., \citep{jain1988algorithms}.
In the present work we divide the training set of data into several clusters
before feeding it into the neural network. Most of the clustering
techniques consider the true value of the data only and do not take account of
the associated errors in the measurements
\citep{Chaudhuri19981307,2004A&A...422.1113Z,Kumar20076084,
2010IJMPD..19.1049B,2012CIDU...118V,2013MNRAS.430..509I}.
In our clustering mechanism we modify the traditional methods in order to
incorporate the errors in the data points. Further, we do not fix the number
of clusters that we wish to make. Rather, given a criteria, we seek if a data point
(in our case a galaxy) can form a cluster
with other data points or not. If not, we consider that the data point as isolated.
Below we describe our clustering mechanism in details.

A data point in the training set have $L$ input patterns (i.e. the
photometric measurements in our case) and one desired output (in
our case the spectroscopic redshift, $z_{\rm spec}$).
Since the photometric redshift ($z_{\rm phot}$) depends more on the difference
of fluxes in different bands rather than the flux in individual bands
we only consider {\it all} possible combinations
of flux differences as $L$ input patterns. This means for a galaxy
with 5 bands photometric measurements we have $L={^5C_2}=10$ differences;
we denote them as $x_i$, $i=1$ to $L$. Each $x_i$ would have
an error associated with them and we add the errors of the corresponding
photometric bands in quadrature to determine error in $x_i$ and call it $\delta x_i$.
Similarly, a cluster (say $C_l$, the $l\:$th cluster) also has $L$ input patterns,
$m_i$ (the average $x_i$ of all cluster members), $L$ dispersions ($\sigma^m_i$) associated
with each pattern and one output pattern, the average redshift of the members
of the cluster, $z_{\rm cl}$. Thus a cluster can be described like a function
$C_l(m_i,\sigma^m_i,z_{\rm cl})$ where $i$
runs from (in our case) 1 to $L={^NC_2}$, if $N$ is the number of photometric
bands for which measurements are available.
Further, $m_i$, $\sigma^m_i$ and $z_{\rm cl}$ are calculated as follows.
Consider that the cluster consists of $N_l$ members/galaxies.
The weighted average of cluster properties, $m_i$,
are calculated as
\begin{equation}
m_i = \mathlarger{{{\mathlarger\sum_{k=1}^{N_l}\cfrac{x_i^k}{(\delta x_i^k)^2}}}\left/
{{\mathlarger\sum_{k=1}^{N_l}\cfrac{1}{(\delta x_i^k)^2}}}\right.}, ~~~~ {\rm for} ~~i=1~~ {\rm to}~~L.
\label{eqn_m_i}
\end{equation}
Here we use the errors for each member as weights to calculate the mean and the index $k$
indicates the particular member. The dispersions in
$m_i$ are obtained from
\begin{equation}
\sigma^m_i = \sum_{k=1}^{N_l} (x_i^k - m_i)^2 / (N_l-1), ~~~ {\rm for}~~ N_l > 1,
\end{equation}
and the average redshift of the cluster is
\begin{equation}
z_{\rm cl} = \sum_{k=1}^{N_l} z_{\rm spec}^k/N_l.
\label{eqn_z_cl}
\end{equation}

We start with a single galaxy
that is not part of any cluster yet and consider
it to be the first member of a cluster, $C_l$.
At this point, the cluster has $m_i=x_i$, $i=1$ to $L$ and $z_{\rm cl}=z_{\rm spec}$.
Since the cluster has only
one member, we assume $\sigma^m_i=0.1m_i$.
Now we search entire data base
to find other members that have similar properties of the cluster. For each
galaxy that has not been included in any existing clusters,
we calculate the {\it input similarity} as
\begin{equation}
\mu_{in}=\mathlarger{\mathlarger\sum_{i=1}^L\left[ 
\dfrac{(x_i-m_i)^2}{\delta x_i\delta x_i+\sigma^m_i\sigma^m_i} \right]}.
\end{equation}
Note that we use both the errors in $x_i$ and dispersion of properties in
cluster members to calculate the probability function $\mu_{in}$.
This number $\mu_{in}$ represents how much the galaxy differs from
the cluster, $C_l$ in terms of input properties and actually is the log
likelihood function excluding some constant terms.
We can understand that if $\mu_{in}$ is less than $L$,
roughly, all $x_i$ including associated errors are within one sigma of the
mean of the cluster. We say that the galaxy
passes the input similarity test with that cluster if $\mu_{in} < \mu^{th}_{in}L$. Here, $ \mu^{th}_{in}$
is some predefined number of order unity which governs the similarity test.
Further, we calculate the {\it output similarity} between a galaxy
and cluster as $z_{\rm cl}-z_{\rm spec}$ with  $z_{\rm cl}$ is the average redshift of the
cluster members and $z_{\rm spec}$ is the spectroscopic redshift of the galaxy
in consideration. Note that we do not consider any error in this case as
spectroscopic redshifts have very small errors. A galaxy is said to pass
the output similarity test if $|z_{cl}-z_{\rm spec}| < \Delta_z$, where $\Delta_z$
is some predefined threshold value. If a galaxy passes both the input
and output similarity tests, it is included in the cluster and we update
the cluster properties using Eqs.~\ref{eqn_m_i}-\ref{eqn_z_cl}.

\begin{figure}
\centerline{\epsfig{figure=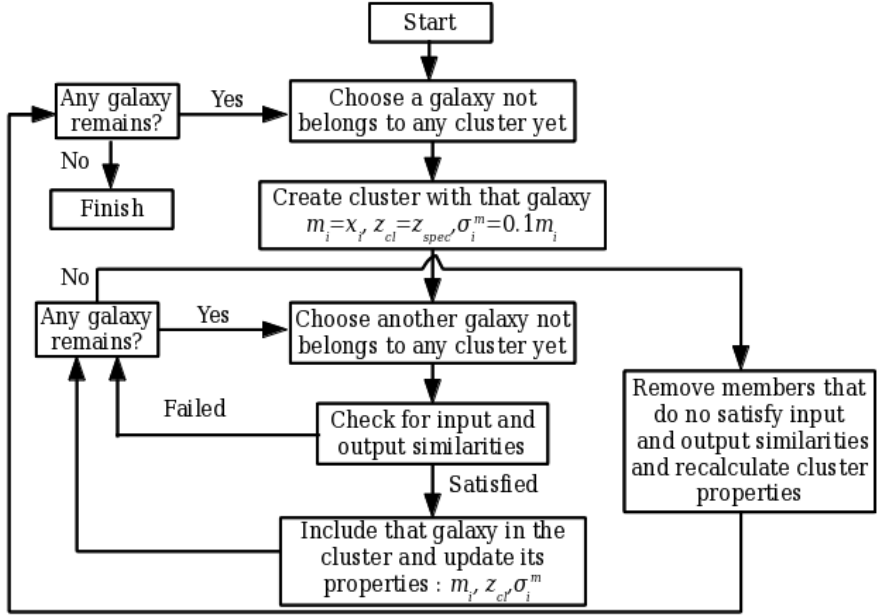,width=12cm,angle=0.}}
\caption[]{The flow chart for creating clusters from the galaxy sample.}
\label{fig_flow_cl}
\end{figure}

We search the entire data set in this process to find all possible
members of the cluster. Note that both $m_i$ and $z_{\rm cl}$ are dynamic
properties of the cluster and are changing as each new member is added to the
cluster. Hence it is possible that for some of the members the input
and output similarity tests may result negative after all members of the
cluster are identified. We check for such members
and remove them from the cluster and recalculate the cluster properties.
We do this iteration only once. When it is finished, we start again with some other
galaxy that is not part of any existing clusters to form a new cluster.
This process is repeated until all the galaxies have been considered
to form a cluster. It may possible that a galaxy does not find
any other galaxy to form cluster and we consider such galaxies
as isolated ones. Moreover, the clusters with number of members
less than 20 are not considered to take part in the neural network
training as an individual cluster due to very small number statistics.
Hence, we try to distribute those
galaxies in existing clusters if they pass the input and output
similarity tests. We check this when all the clusters are already
been identified.
The entire algorithm is shown by a flow chart in Fig.~\ref{fig_flow_cl} 

Note that all galaxies in the training set do not necessarily be part of some cluster.
Both $\Delta_z$ and $\mu^{th}_{in}$ determine the number of clusters
and the size and dispersion of the individual clusters as well as the number of galaxies
that belong to all the clusters. For example, using $\Delta_z=0.01$
and $\mu^{th}_{in}=1.0$ produce 251 clusters having more than 20
members and only 9330 galaxies out of 20809 galaxies form those clusters.
On the other hand, making $\mu^{th}_{in}=1.2$ creates 262 clusters including
total 12298 galaxies. We use these clusters to train separate neural networks.

\section{Back propagation Neural Network}
\label{sec_bnn}
Artificial neural network has widely been used in different
fields of machine learning
\citep{1994ApJ...426..340G,1994VA.....38..293G,1997A&A...322..933G,1996Ap&SS.239..361M,
1998MNRAS.295..312S,2008Ap&SS.315..201B,2009NewA...14..649B}.
Here we use neural network with back propagation learning to estimate photometric
redshift from the fluxes at different bands. As the name suggests,
the error in the output node propagates backward to update weights
at different layers of network. This is also called supervised learning methods
\citep[see][for more details]{bishop1995neural}. The first layer of our network
consists of $L$ nodes, the $L$ colors. Note that we rescale
all values between 0 and $1$ in the first layer of nodes. This node
is connected to an intermediate hidden layer of $p$ nodes
where $p$ is equal to the closest integer of $\sqrt{L}$.
We find this is the optimised number of nodes. Increasing this
would not improved much but takes much more computational
time, where as reducing would lead to poorer results.
The first and hidden layers are connected by weight parameters
$V_{ij}$ where $j$ runs from 1 to $p$ and $i$ ranges from 1
to $L$. The hidden layer is connected to output layer by weight
parameters $W_j$, $j=1$ to $p$. We use
$f(x)=[1-\exp(-\lambda x)]/[1+\exp(-\lambda x)] $
as the activation function with $\lambda$ is some parameter that governs
the slope of the activation function.
The advantage of taking this activation function is that it is nonlinear,
continuous, differentiable and bounded between -1 to 1 and such functions
provide excellent result in gradient base method
\citep{NIPS1993_874}. 
We start with $\lambda=1.5$ and
gradually lower its value as the iteration increases. This ensures to
reach much closer to the global minimum.
In the feed-forward neural network method the output photometric
redshifts are obtained from \citep{bishop1995neural},
\begin{equation}
H_j=\sum_{i=1}^{L}x_iV_{ij}
\label{eqn_hj}
\end{equation}
\begin{equation}
z_{\rm phot}=\sum_{j=1}^{p} f(H_j)W_j.
\label{eqn_zphot}
\end{equation}

The weights are updated using back propagation
gradient descent method as,
\begin{eqnarray}
\eth &=& (z_{\rm spec} - z_{\rm phot})f^\prime(z_{\rm phot}) \\
\delta W_j &=& \alpha \eth f(H_j) \\
\delta V_{ij} &=& \alpha \eth W_j f^\prime(H_j) x_i.
\end{eqnarray}
Here $\delta W_j$ and $\delta V_{ij}$ are the increments in the weights
and $\alpha$ is the learning parameter.

Note that we use separate networks for each clusters having member
greater than 20 and another separate network for the entire galaxy sample.
For clusters we use maximum 2000 iterations and for the whole sample
it is 1,00,000 and take the best weight factors that minimize
the error function, $\langle (z_{{\rm spec}}-z_{{\rm phot}})^2\rangle^{1/2}$.

\section{Photometric redshift using CuBAN$z$}
\label{sec_result}
\begin{figure}
\centerline{\epsfig{figure=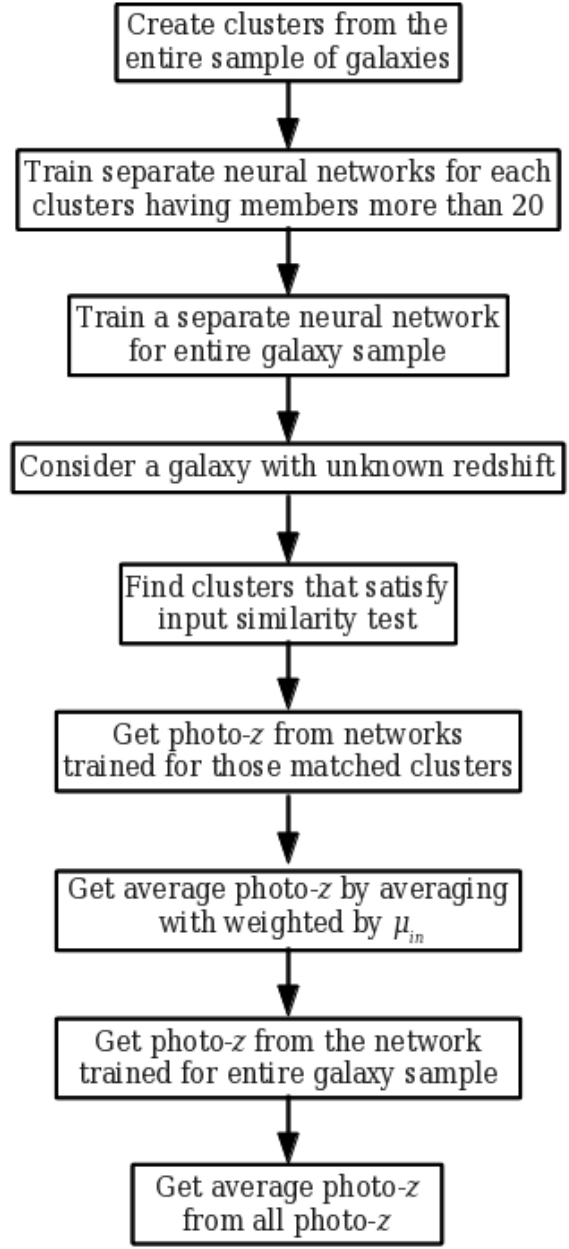,width=8cm,angle=0.}}
\caption[]{The flow chart for getting photometric redshifts in CuBAN$z$.}
\label{fig_flow_cubanz}
\end{figure}

CuBAN$z$ is the name of our photometric redshift estimator. The name
is derived from {\bf C}l{\bf u}stering aided {\bf Ba}ck propagation {\bf N}eural
network photo-{\bf $z$} estimator. As the name suggests we use
clustering of training sources (as already described in 
section~\ref{sec_cluster}) first and then use back propagation
neural network (described in the previous section) to
train the networks and use them to get photometric redshifts
for unknown sources. For an unknown source we first look for
clusters that satisfy the input similarity test i.e.
$\mu_{in} < \mu^{th}_{in}L$, then use the networks trained for those
clusters to estimate the photometric redshifts and corresponding
errors. If a galaxy has more than one cluster satisfying the
input similarity test, we use weighted average to estimate
the redshift of that galaxy. The values of $\mu_{in}$ are used
as the inverse of the weights.
Note that a smaller value of $\mu_{in}$ implies a better
match with the corresponding cluster and hence get more weighting in determining the
average photometric redshift. Further, we use the network trained with the whole sample
to estimate the redshift for galaxies that do not pass input similarity
test with any training set clusters\footnote{We call sources that pass similarity
tests as clustered sources and others as unclustered sources.}.
Since each cluster having members
on average of between 50 to 100, we also provide a weighted average
of redshift calculated using the clustered networks and the whole
sample to remove any bias due to small number. We use the uncertainty
in the redshift as the weights in this case.
In Fig.~\ref{fig_flow_cubanz} we show the block diagram
that CuBAN$z$ follows.

The error in the estimate of the photometric redshift is a major concern
and we deal with that quite rigorously here. From Eqn.~\ref{eqn_hj}
and \ref{eqn_zphot}, we can say that the photometric redshift derived
using neural network is a function of $x_i$, $W_j$ and $V_{ij}$, i.e. 
\begin{equation}
z_{\rm phot }=z_{\rm phot}(x_i,W_j,V_{ij}).
\end{equation}
Therefore, the error in $z_{\rm phot }$ can be thought of due to two parts,
one due to errors
in the $x_i$ i.e. in the flux measurements, and secondly uncertainty
in the values of $W_j$, $V_{ij}$ that are obtained through the training.
The error due to photometric uncertainty can be calculated by taking the
derivative of the Eqn.~\ref{eqn_zphot} keeping $W_j,~V_{ij}$ as constants i.e.
\begin{eqnarray}
\delta z_{\rm phot}& = &\sum_{j=1}^{p} \delta f(H_j)~~W_j = \sum_{j=1}^{p}
\frac{\partial f}{\partial H_j}\delta H_j ~~ W_j \\
\delta H_j & = & \sum_{i=1}^{L} \delta x_i V_{ij}
\end{eqnarray}
However, the error due to uncertainty in the weights can not be calculated
in this way. Since there is no straight forward measure of the uncertainty of
the weights we take the residual rms in the training set as a proxy
for the total uncertainty in redshift due to uncertainty in all the weights
combined. Then we add these two errors in quadrature to get the final
uncertainty in the value of photometric redshift.

\begin{figure}
\centerline{\epsfig{figure=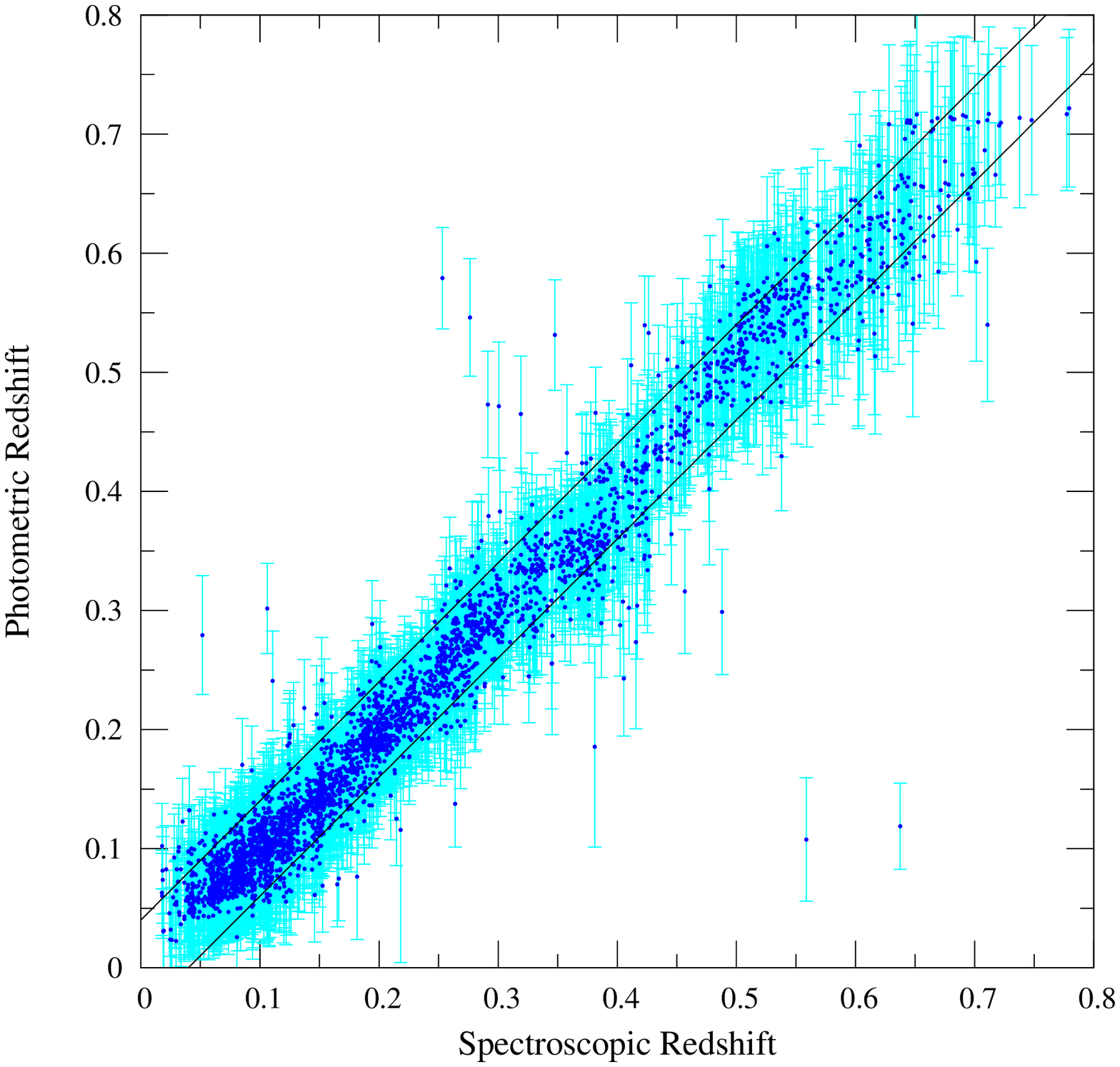,width=10cm,angle=-90.}}
\centerline{\epsfig{figure=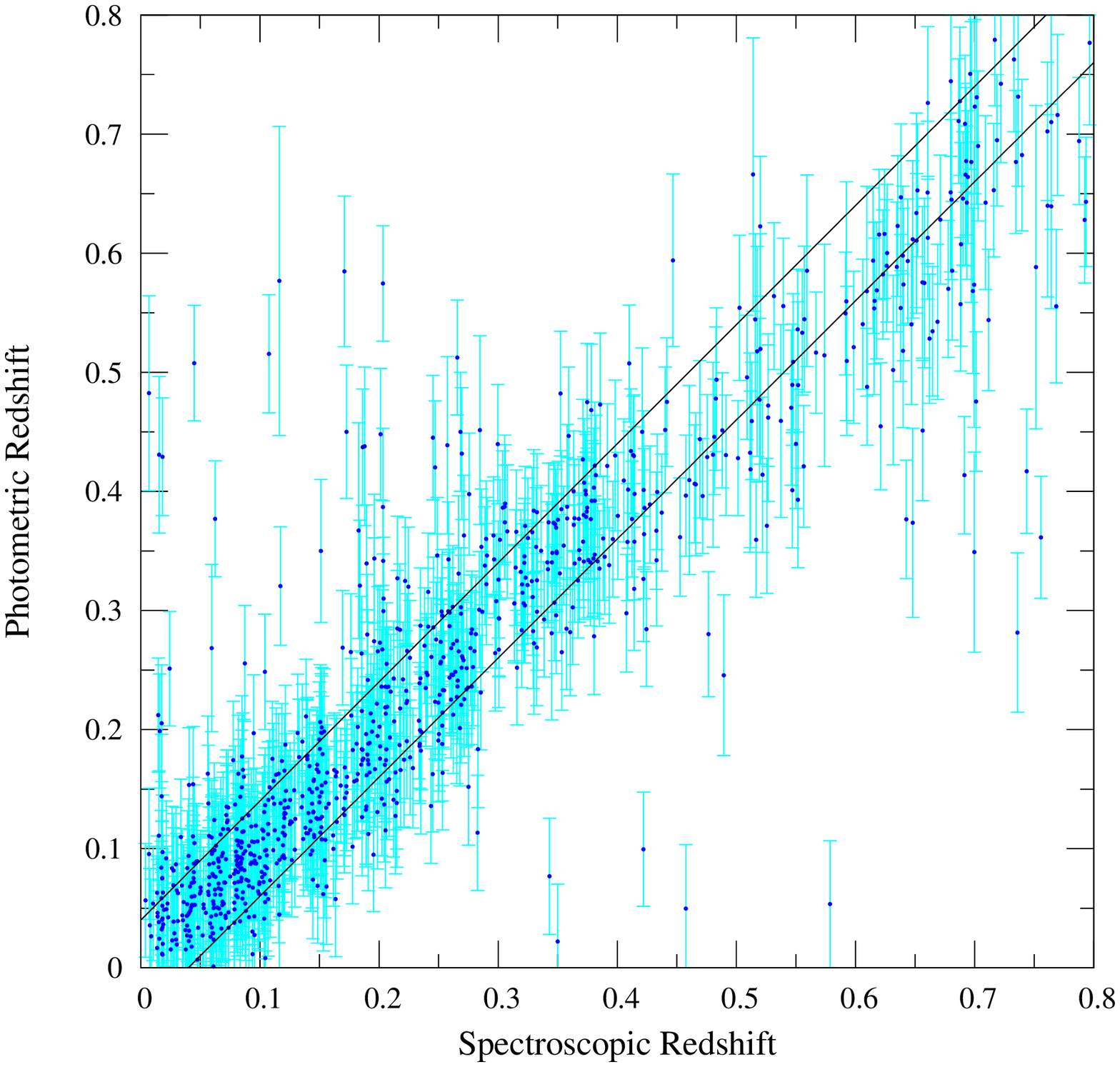,width=10cm,angle=-90.}}
\caption[]{The spectroscopic redshifts vs photometric redshifts as obtained
from CuBAN$z$ for the test set galaxies. Top panel is for galaxies that
pass input similarity test with some cluster and bottom panel is for other galaxies that
do not pass the similarity test with any cluster. The parallel lines represent $z \pm 0.04$. }
\label{fig_test_1.2}
\end{figure}

We now show our results using the code CuBAN$z$ on the SDSS stripe 82
catalog. Fig.~\ref{fig_test_1.2} shows the prediction of photometric
redshift by CuBAN$z$ against spectroscopic redshift for our 4213 test
galaxies. Note that our training set consists of 20809 galaxies.
Here we assume $\Delta_z=0.01$ and $\mu^{th}_{in}=1.2$. These criteria produce
262 clusters with more than 20 members. All these clusters contain
a total of 12298 galaxies which is $\sim 60\%$ of total training set.
For the clusters the rms errors
(i.e. $\langle (z_{{\rm spec}}-z_{{\rm phot}})^2\rangle^{1/2} $)
are of the order of few times $10^{-3}$
and for the whole sample final rms error is 0.048 on the training
set itself. For the test set consisting of 4312 galaxies, we find
3406 galaxies pass the input similarity test with one or more training set
clusters and rest 906 galaxies do not pass similarity test
with any cluster of the training set. For the 3406 clustered sources, the rms
error is as low as 0.034 where as for the rest of the sample is 0.097 making a total
of 0.054 for the entire sample. Note that in the training set there are
very few galaxies with $z >0.7$ (around 300). Hence we should not consider
anything beyond $z=0.7$. If we restrict ourself with such criterion
the rms error for the entire sample reduces to 0.045. In compare to
the existing ANNz code the same set of training and testing set produces
rms error of 0.055.
We summarise our results in Table~\ref{tab_results}.

\begin{table}
\begin{center}
\begin{tabular}{|l|c|c|} \hline
Galaxy sample & No. of galaxies & rms error \\ \hline
Training set: All &  20809   & 0.048 \\ \hline
Testing set: Clustered & 3406 & 0.034 \\ \hline
Testing set: non-clustered & 906 & 0.097 \\ \hline
Testing set: All     & 4312 & 0.054 \\ \hline
Testing set: $z\le 7.0$ & 4005 & 0.045 \\ \hline
\end{tabular}
\caption[]{The rms values for different galaxy samples.}
\label{tab_results}
\end{center}
\end{table}

Hence, CuBAN$z$ provides very accurate redshift estimation from
photometric measurements especially sources
that match with existing clusters (i.e. pass the input similarity test)
of the training set. This is clearly
reflected in the top panel of Fig.~\ref{fig_test_1.2} as well. 
In Fig.~\ref{fig_test_1.2} we show photo-$z$ estimated from CuBAN$z$
against the spectroscopic redshift for both clustered (top panel)
and unclustered (bottom) sources/galaxies in the testing set
along with the estimated uncertainty.
Only a handful of outliers is present there in the top panel.
Most of the points along with the estimated errors lie well
within the $z\pm 0.04$ limits as shown
by the two parallel lines\footnote{$45^\circ$ parallel lines in
Fig.~\ref{fig_test_1.2} and in other figures corresponds to $z\pm 0.04$ limits.}.
On the other hand the unclustered sources
show little more scatter (bottom panel of Fig.~\ref{fig_test_1.2}) compare
to the clustered sources. This is also reflected naturally on the estimated
errors as we deal with them properly. The errors in
the estimated photometric redshifts for the
clustered sources are on average lower compared to the errors estimated for
the rest of the sources. 

\begin{figure}
\centerline{\epsfig{figure=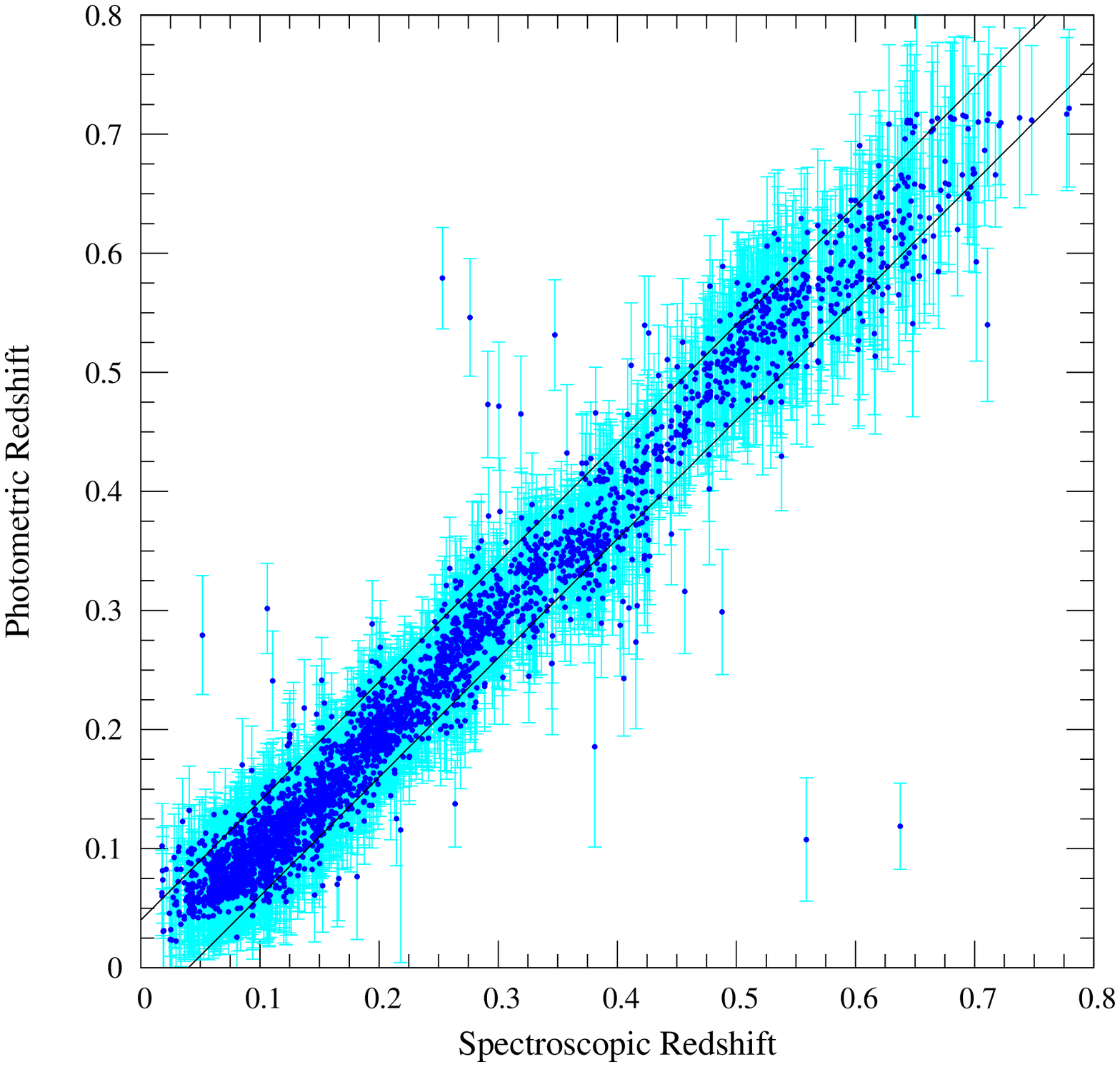,width=5.0cm,angle=-90.}
\epsfig{figure=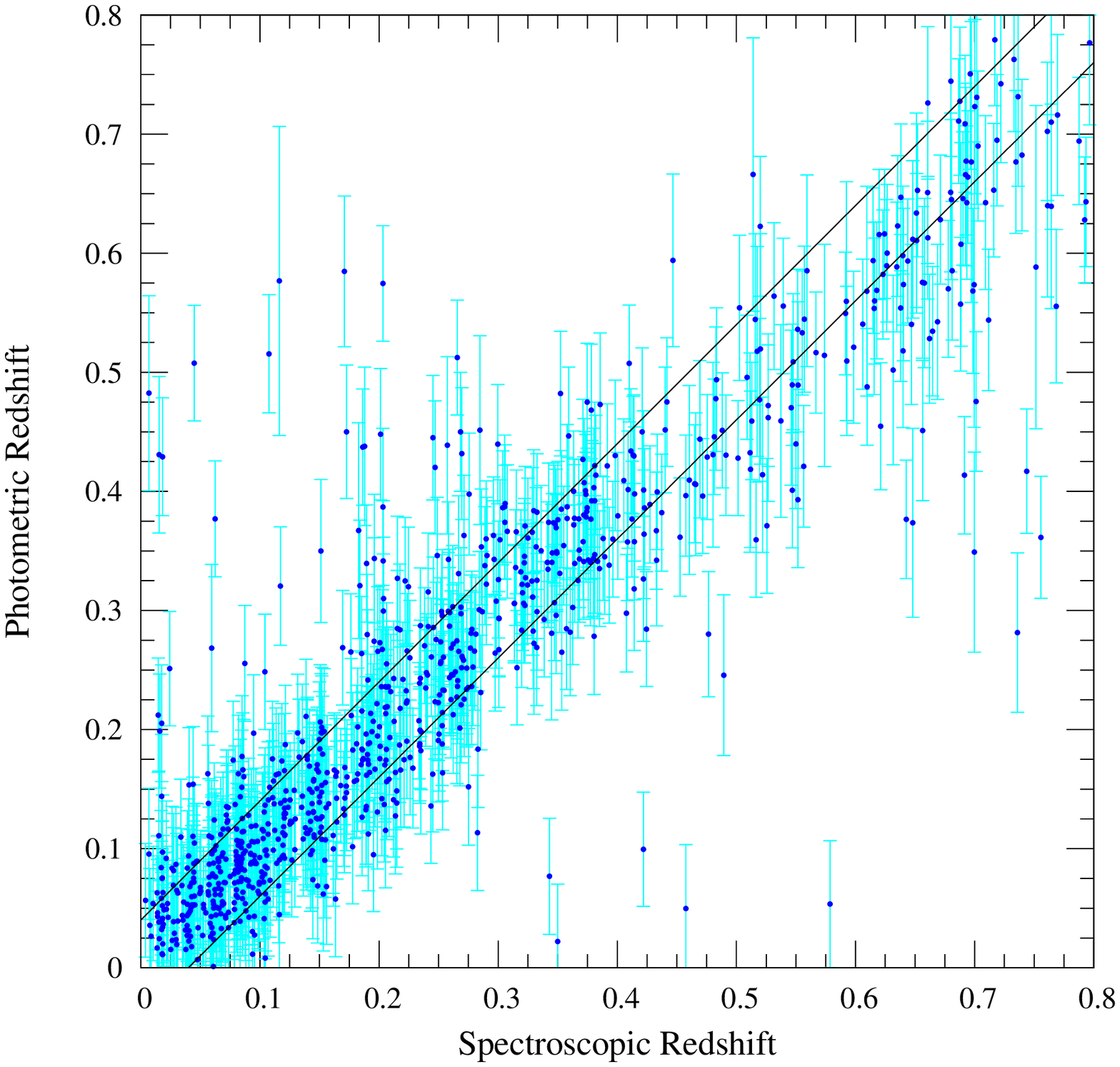,width=5.0cm,angle=-90.}
\epsfig{figure=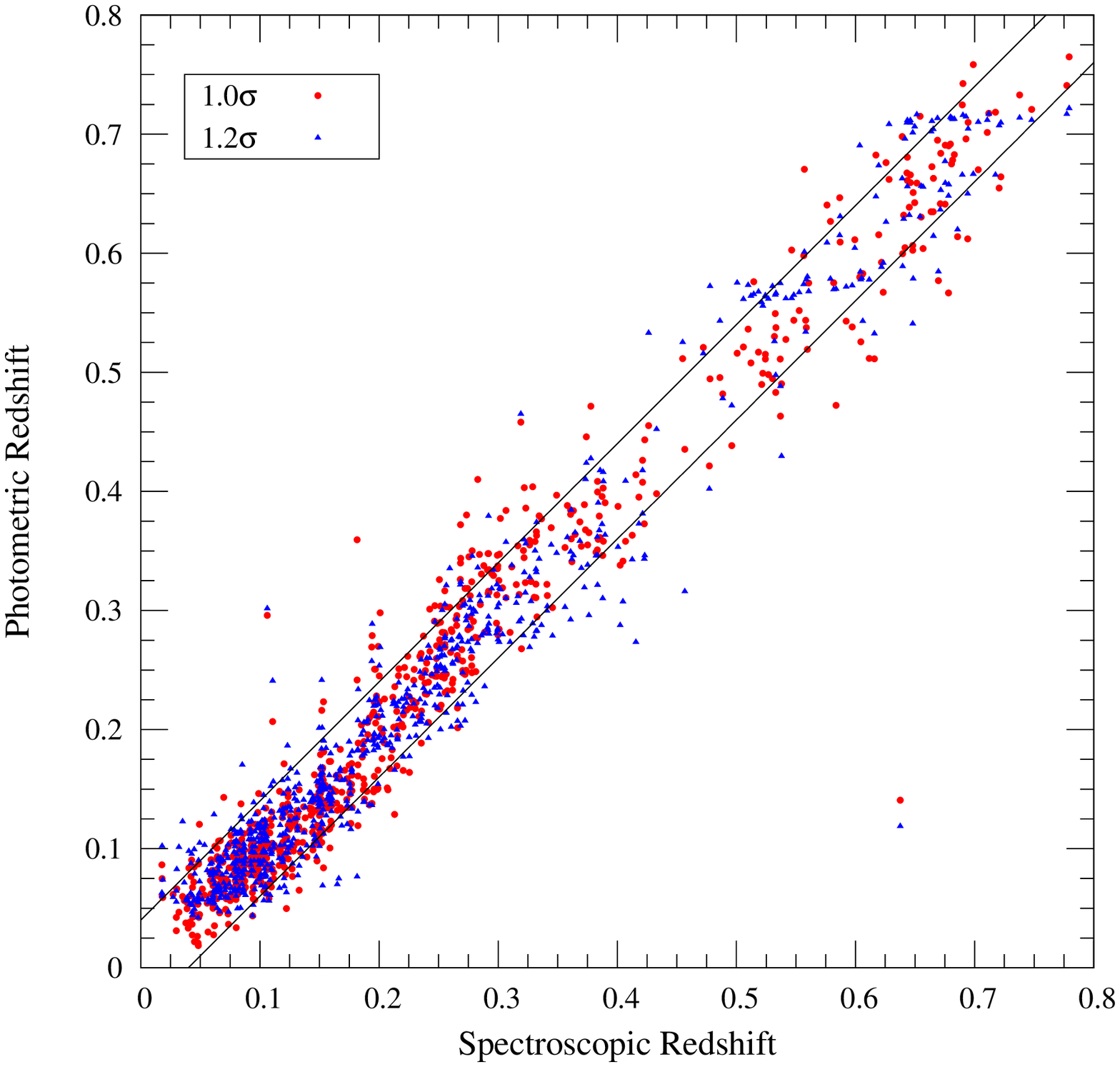,width=5.0cm,angle=-90.}
}
\centerline{\epsfig{figure=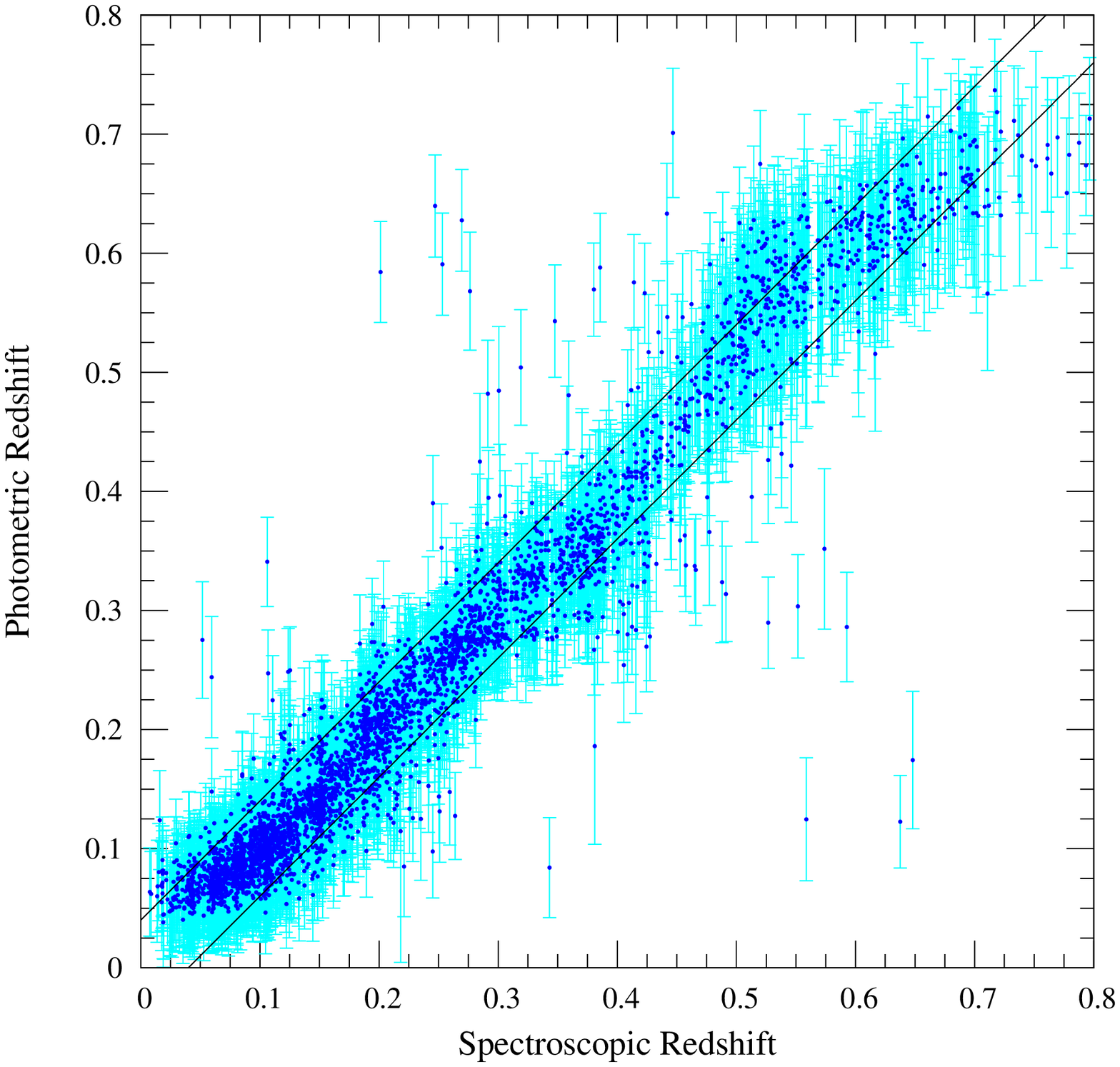,width=5.0cm,angle=-90.}
\epsfig{figure=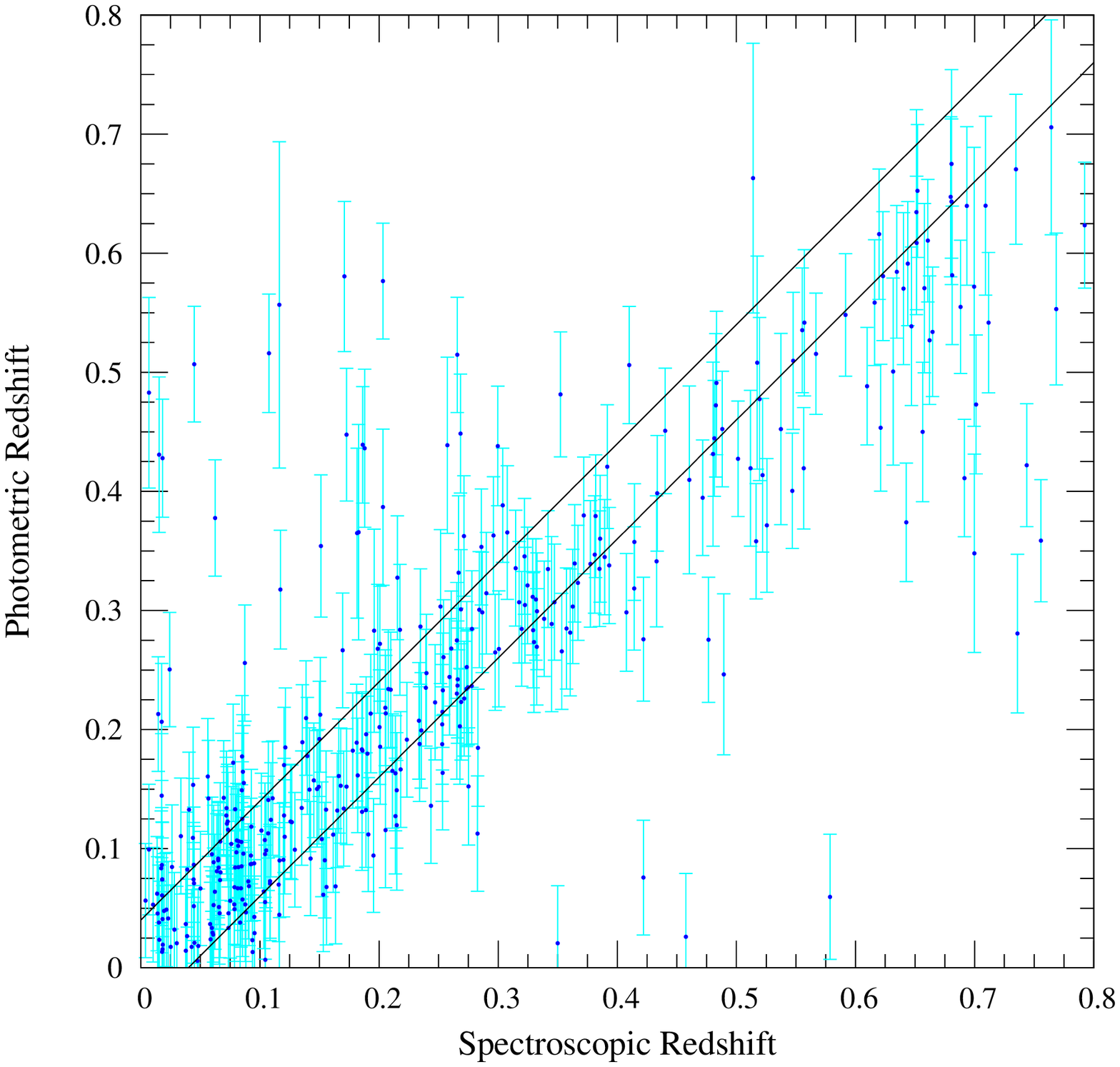,width=5.0cm,angle=-90.}
\epsfig{figure=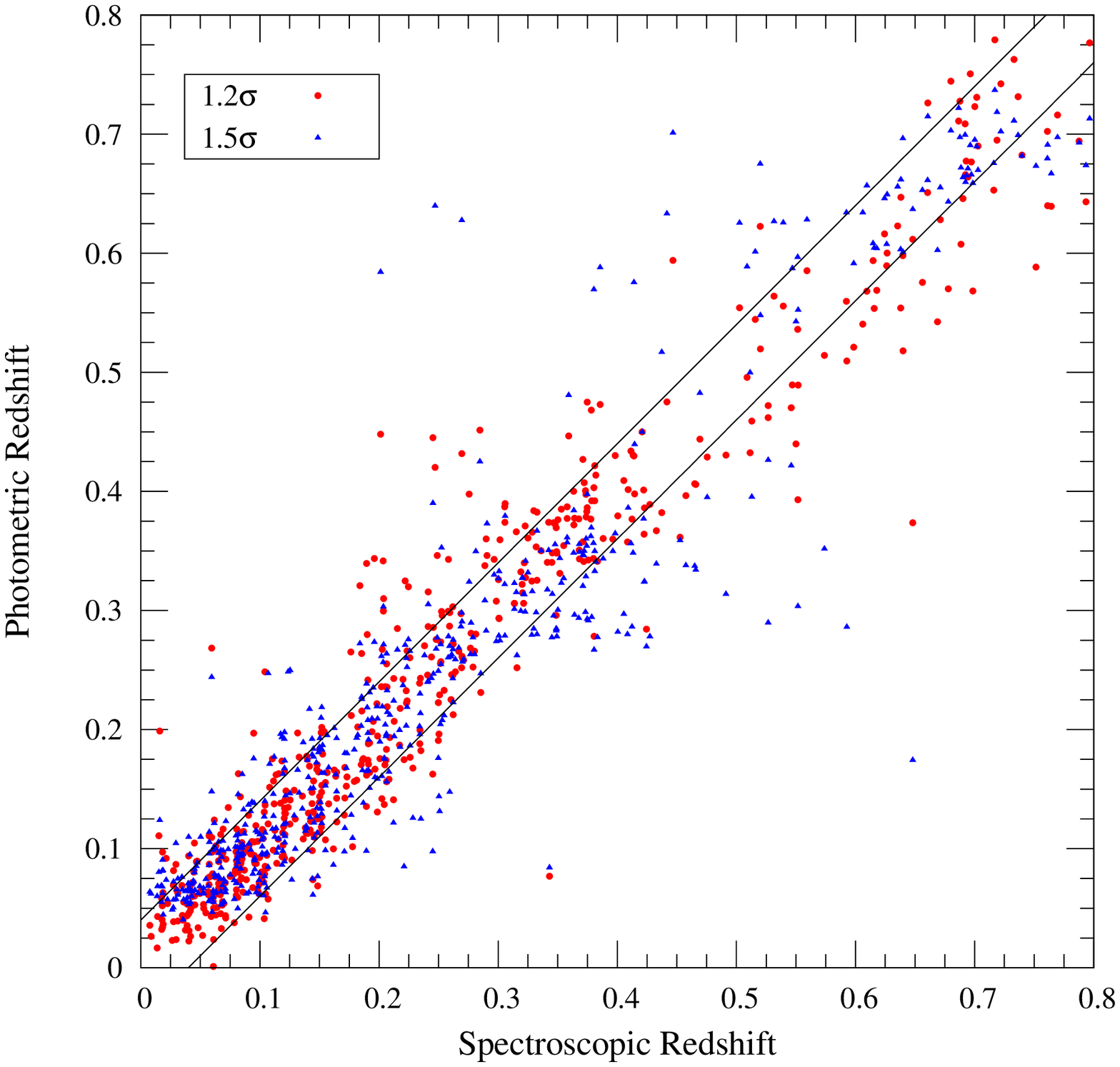,width=5.0cm,angle=-90.}}
\caption[]{The spectroscopic redshifts vs photometric redshifts obtained from
CuBAN$z$ for the test set galaxies. Top panels for $\mu_{in}^{th}=1.0$
and bottom panels for $\mu_{in}^{th}=1.5$. The left panels are for clustered
sources where as the middle panels are for unclustered sources. The right
panels show the difference of redshift for the sources when calculated
using the entire training set (red bullets) and the clustered training set (blue triangles).}
\label{fig_test_1.0_1.5}
\end{figure}

Thus the clustering is an important leap in predicting the photometric redshift.
The properties of clusters depend on the two parameters, $\Delta_z$ and
$\mu_{in}^{th}$. In Fig.~\ref{fig_test_1.0_1.5} we show the effect of
cluster properties on the estimated redshifts. In the top panel
we change $\mu_{in}^{th}=1.0$. This leads to form 251 clusters
with 9330 galaxies (only 45\% of total sample). In case of the testing set,
only 2684 sources pass the input similarity test with training set clusters.
In this case the rms errors for the clustered and unclustered
sources are 0.034 and 0.077 respectively making total rms 0.54.
Thus it does not make any comprehensible difference in predicted
photo-$z$. The rms for unclustered sources is reduced as the number of
unclustered sources has increased but CuBAN$z$ is still predicting
very good redshift for them using the network trained by the entire galaxy sample.
The difference between 
$\mu_{in}^{th}=1.0$ and 1.2 is shown in top right panel of 
Fig.~\ref{fig_test_1.0_1.5} where we show the photometric redshift estimated
for the sources that belongs to some cluster when $\mu_{in}^{th}=1.2$ but
do not fall in any cluster when $\mu_{in}^{th}=1.0$. These are the sources
for which the photo-$z$ are  calculated in two different methods; when
$\mu_{in}^{th}=1.0$ photo-$z$s are obtained from the network that was
trained using the entire galaxy sample where as in other case the 
photo-$z$s are obtained using networks for the clustered sources.
 It is clear from the
figure that there is not much difference in this case. However, assuming
$\mu_{in}^{th}=1.5$ shows prominent effect. With this assumption, more
sources become part of clusters, $15641$ in case of training set and 3938 for
the testing set. Again, the difference is clear in the bottom right panel
of Fig.~\ref{fig_test_1.0_1.5} where the scatter in predicted photometric
redshifts from the two different networks (cluster networks and whole sample network)
is obvious. In this case the clustering is producing clusters
having member with a larger dispersion in photometric
properties as we have relaxed the criteria of being part of a cluster
and that is reflected in
the predicted photometric redshifts.
Thus we find $\mu_{in}^{th}=1.2$ provides optimised results for this
data set, and we use it to do the further analysis of getting
photometric redshift for the entire stripe 82 catalog.

\begin{figure}
\centerline{\epsfig{figure=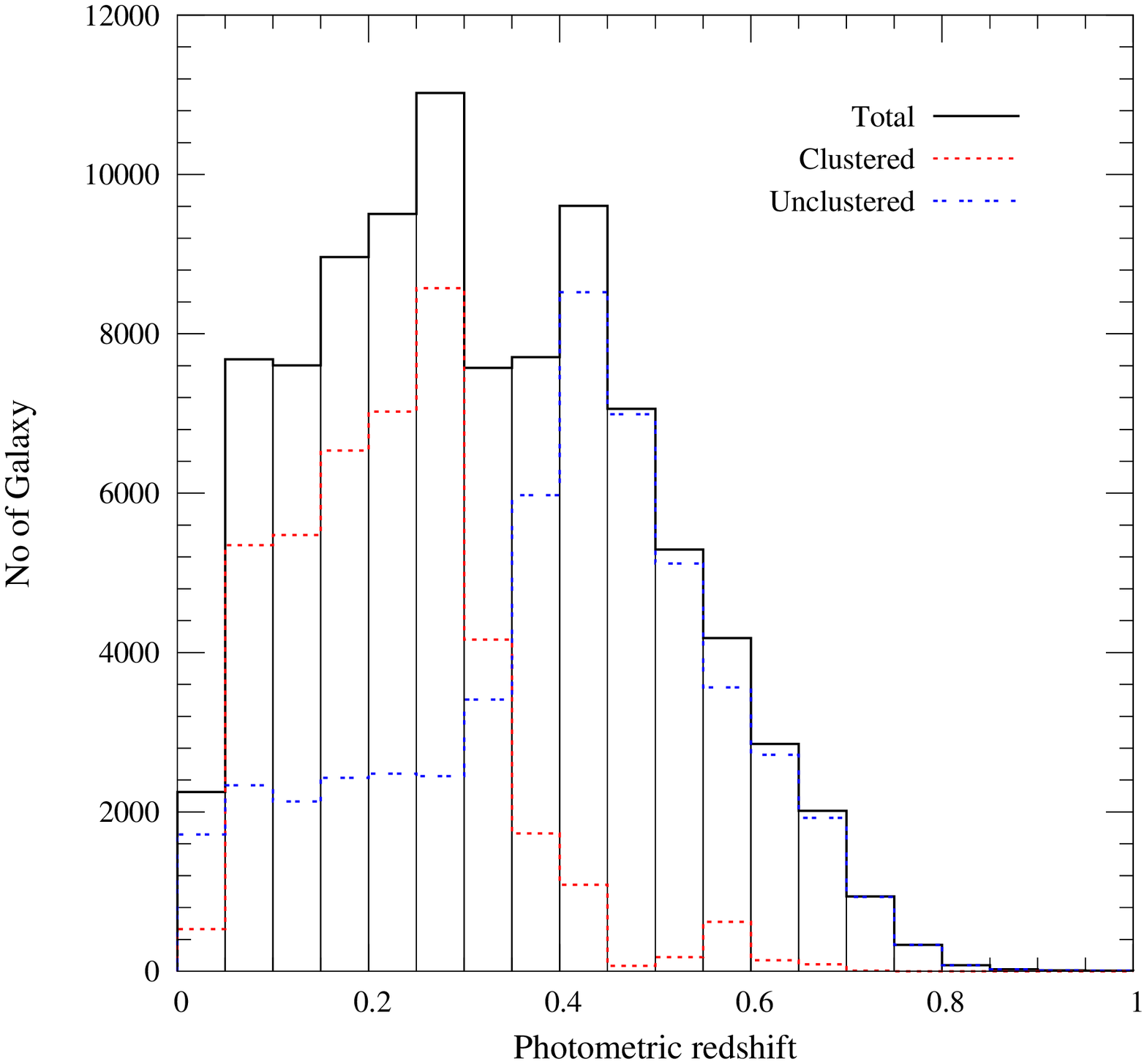,width=10cm,angle=-90.}}
\caption[]{Photometric redshift distribution of half of the SDSS Stripe 82 catalog
as obtained using CuBAN$z$.
 }
\label{fig_s82_hist}
\end{figure}

In Fig.~\ref{fig_s82_hist} we show the distribution of photo-$z$
for the half of the stripe 82 catalog consisting of 94718 galaxies.
Note that we do not put any
flux cut off for these sample (see section~\ref{sec_data}). As expected the redshift distribution
shows a peak at $z=0.3-0.4$ as the effective survey volume increases
with increasing redshift.
However, after that the number of sources decreases due to decreasing
flux of the sources for larger distance. Hence CuBAN$z$ is providing
a good estimation of the redshift. Further, it is also obvious
from the figure that the sources that pass the similarity
test decrease rapidly at higher redshift as the number of training
sources/clusters decreases.

\begin{figure}
\centerline{\epsfig{figure=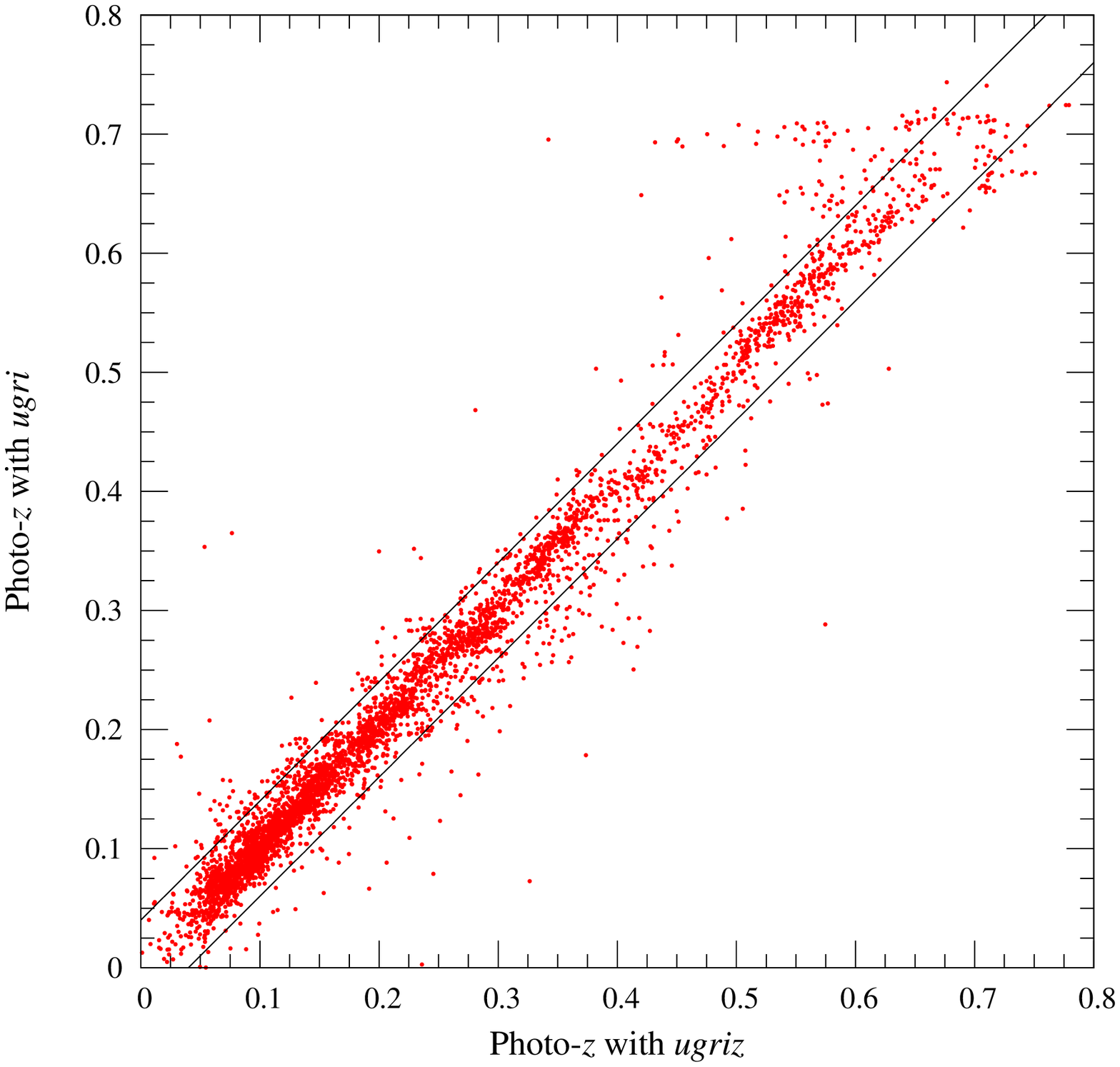,width=10cm,angle=-90.}}
\caption[]{Comparison of photometric redshift estimated by CuBAN$z$ using 5 bands ($ugriz$)
and 4 bands ($ugri$) photometric data.
 }
\label{fig_band45_comp}
\end{figure}

Finally, we show that CuBAN$z$ provides reasonable results even if we have
measurements for lesser number of photometric bands. This is resulted from the use
of colors rather than the individual fluxes in training the
neural networks. Fig.~\ref{fig_band45_comp} compares the photometric
redshift estimated for 5 bands and 4 bands photometry of
the same data. No obvious mismatch is observed except perhaps
at $z>0.7$ where we have very small number of galaxies in training set
itself. Further, CuBAN$z$ also provides better results even if there
are less number of data for training. We randomly choose 2000 galaxies to form
a new training set. With this set the rms for 4312 test set galaxies in CuBAN$z$
is 0.06 where as ANNz provides rms error of 0.10.

\section{User Manual of CuBAN$z$}
\label{sec_code}
CuBAN$z$ is freely available from \url{https://goo.gl/fpk90V} or
\url{https://sites.google.com/site/sscubanz/} under gnu public license agreement.
We provide entire source code written in C which is very easy to compile,
run and even modify. It can be run on any machine having standard C compiler. In the package, we provide
a header file, {\it cnn.h}, a text file containing the names of catalog files,
{\it file\_names.txt}, the source code, {\it cubanz.c} and a {\it README} file for general
instructions. In the {\it cnn.h} file there are two lines as

\texttt{\#define no\_of\_galaxy 20809 }

\texttt{\#define no\_of\_bands 5}\\
\texttt{no\_of\_galaxy} is the number of sources available for the training with
spectroscopic redshift and  \texttt{no\_of\_bands} is the number of photometric
bands data that the training set has. One needs to modify these values according
to the requirement. The catalog names and output file names
are to be mentioned in the {\it file\_names.txt} file in the following order:\\
1. Name of file containing the training set \\
2. Name of file containing the test data set \\
3. File name where output for test data would be written \\
4. File name of catalog for which you wish to get photo-$z$ \\
5. File name where output for given catalog (4) would be written. \\
The training and testing catalog should have all photometric measurements first
and then their corresponding errors. The first $N$ columns should have
the apparent magnitudes of $N$ photometric bands and next $N$ columns
should contain the corresponding errors in magnitude. The last column should have the
spectroscopic redshift. The catalog for which one wishes to get the photo-$z$
should have the data in same order except the redshift information of course.

The source code can be compile with any standard C compiler say, cc, as
{\it cc cubanz.c -lm -o cubanz.o}. This will produce the executable {\it
cubanz.o}. Running the executable will give the desire photo-$z$.
The output file for the test data would contain the following columns:
(1)~serial number, (2)~spectroscopic redshift, (3)~photometric redshift from
clusters (if the source does not pass input similarity test with any
training cluster it would be -10),
(4)~error in column 3, (5)~value of best $\mu_{in}$ (for unclustered
source it is 100), (6)~cluster tag (1 for clustered source, -1 otherwise),
(7)~number of clusters that pass input similarity test with the source,
(8)~photometric redshift using neural
network with whole sample, (9)~error in column 8, (10)~weighted
average photo-$z$ from column~3 and column~8, (11)~error in
column 10. The output for the actual catalog with unknown
redshift will have all the columns except spectroscopic redshift
of course.
 Note that
it will also produce a log file with name {\it cnn\_out.log} that contains
every details of what the code is doing.

\section{Discussions and Conclusions}
\label{sec_cd}
We have introduced a new photometric redshift estimator, CuBAN$z$, that
provides a much better photo-$z$ compared to the existing ones.
The code is publicly available and very simple to use. It can be run
in any machine having standard C compiler.
It uses back propagation neural networks
clubbed with clustering of training sources with known photometric
broad band fluxes and spectroscopic redshifts. The clustering technique
enables us to get a better estimate of photometric redshifts particularly for
galaxies that fall under clusters. In particular the rms residue in the
testing set is as low as 0.03 for a wide redshift range of $z \le 0.7$
compare to existing ANNz code that gives 0.055 on the same data set.
Moreover, we provide much better estimate on the uncertainty in the
redshift estimator considering the uncertainty in the weight
factors of the trained neural networks. We hope that it will be very
useful to the astronomy community given the existing large photometric
data as well as large upcoming photometric surveys. The present version
of the code is very simple and we are in the process of making it
more flexible as well as user friendly.
 
\section*{Acknowlegdements}
SSP thanks Department of Science and Technology (DST), India for the INSPIRE
fellowship. We thank an anonymous referee for useful comments that have helped
in improving our paper.
\bibliographystyle{elsarticle-harv}

\end{document}